\documentclass[twocolumn]{aastex62}

\def\kms{\ifmmode{\rm km\thinspace s^{-1}}\else km\thinspace s$^{-1}$\fi}
\def\ms{\ifmmode{\rm m\thinspace s^{-1}}\else m\thinspace s$^{-1}$\fi}

\graphicspath{{./}{figures/}}

\received{\today}
\accepted{19 Jan, 2021}
\submitjournal{\it{Frontiers - Space Science and Astronomy - Exoplanets: The Effect of Stellar Multiplicity on Exoplanetary Systems}}

\shorttitle{Speckle Interferometry}
\shortauthors{Howell et al.}

\begin{document}

\title{The NASA High-Resolution Speckle Interferometric Imaging Program: \\ Validation and Characterization of Exoplanets and Their Stellar Hosts}


%
\correspondingauthor{Steve B. Howell}
\email{steve.b.howell@nasa.gov}

\author[0000-0002-2532-2853]{Steve~B.~Howell}
\affil{NASA Ames Research Center, Moffett Field, CA 94035, USA}

\author[0000-0003-1038-9702]{Nicholas~J.~Scott}
\affil{NASA Ames Research Center, Moffett Field, CA 94035, USA}

\author[0000-0001-7233-7508]{Rachel A. Matson}
\affiliation{U.S. Naval Observatory, 3450 Massachusetts Avenue NW, Washington, D.C. 20392, USA}

\author[0000-0002-0885-7215]{Mark~E.~Everett}
\affiliation{NSF’s National Optical-Infrared Astronomy Research Laboratory, 950 N. Cherry Ave., Tucson, AZ 85719, USA}

\author[0000-0001-9800-6248]{Elise Furlan}
\affiliation{NASA Exoplanet Science Institute, Caltech/IPAC, Mail Code 100-22, 1200 E. California Blvd.,
Pasadena, CA 91125, USA}

\author[0000-0003-2519-6161]{Crystal~L.~Gnilka}
\affil{NASA Ames Research Center, Moffett Field, CA 94035, USA}

\author[0000-0002-5741-3047]{David R. Ciardi}
\affiliation{NASA Exoplanet Science Institute, Caltech/IPAC, Mail Code 100-22, 1200 E. California Blvd.,
Pasadena, CA 91125, USA}

\author[0000-0002-9903-9911]{Kathryn V. Lester}
\affil{NASA Ames Research Center, Moffett Field, CA 94035, USA}





\begin{abstract} 
Starting in 2008, NASA has provided the exoplanet community an observational program aimed at obtaining the highest resolution imaging available as part of its mission to validate and characterize exoplanets, as well as their stellar environments, in search of life in the universe. Our current program uses speckle interferometry in the optical (320-1000 nm) with new instruments on the 3.5-m WIYN and both 8-m Gemini telescopes. Starting with Kepler and K2 follow-up, we now support TESS and other space- and ground-based exoplanet related discovery and characterization projects. 
The importance of high-resolution imaging for exoplanet research comes via identification of nearby stellar companions that can dilute the transit signal and confound derived exoplanet and stellar parameters. Our observations therefore provide crucial information allowing accurate planet and stellar properties to be determined.
Our community program obtains high-resolution imagery, reduces the data, and provides all final data products, without any exclusive use period, to the community via the Exoplanet Follow-Up Observation Program (ExoFOP) website maintained by the NASA Exoplanet Science Institute. This paper describes the need for high-resolution imaging and gives details of the speckle imaging program, highlighting some of the major scientific discoveries made along the way. 
\end{abstract}



\section{Introduction}
The study of exoplanets is one of the most important topics in astrophysics today. Starting over a decade ago, in support of the NASA Kepler mission, a program providing follow-up observations began. It became clear as Kepler was nearing launch, that the 4 arcsec pixels \citep{Borucki2010} as well as the many possible confounding events which could imitate exoplanet transit events
(e.g., \citealt{Brown2011, Santerne2013}) would require follow-up observations from ground-based telescopes in order to validate and characterize any discovered transit candidates. In addition, for transit observations it is crucial to know the stellar properties well, since the planet radius depends directly on the stellar radius. Also, given the relatively large pixels and multi-pixel photometric apertures, it is possible that more than one star is measured, and thus the transit measurement becomes even more uncertain or unreliable.

To support exoplanet discovery, spectroscopic follow-up observations consisted of medium- and high-resolution work using reconnaissance spectra
at the start and then large telescope efforts once specific validation steps were passed \citep{Furlan2018}. Likewise, imaging observations were performed ranging from standard native seeing CCD imaging and lucky imaging to high-resolution observations \citep{Furlan2017}. These latter consisted of both Infrared Adaptive Optics (IR/AO) observations using Lick, Palomar, and Keck and optical speckle interferometric imaging using WIYN and Gemini telescopes.

\begin{figure*}[t]
\centering
\includegraphics[scale=0.55,keepaspectratio=true]{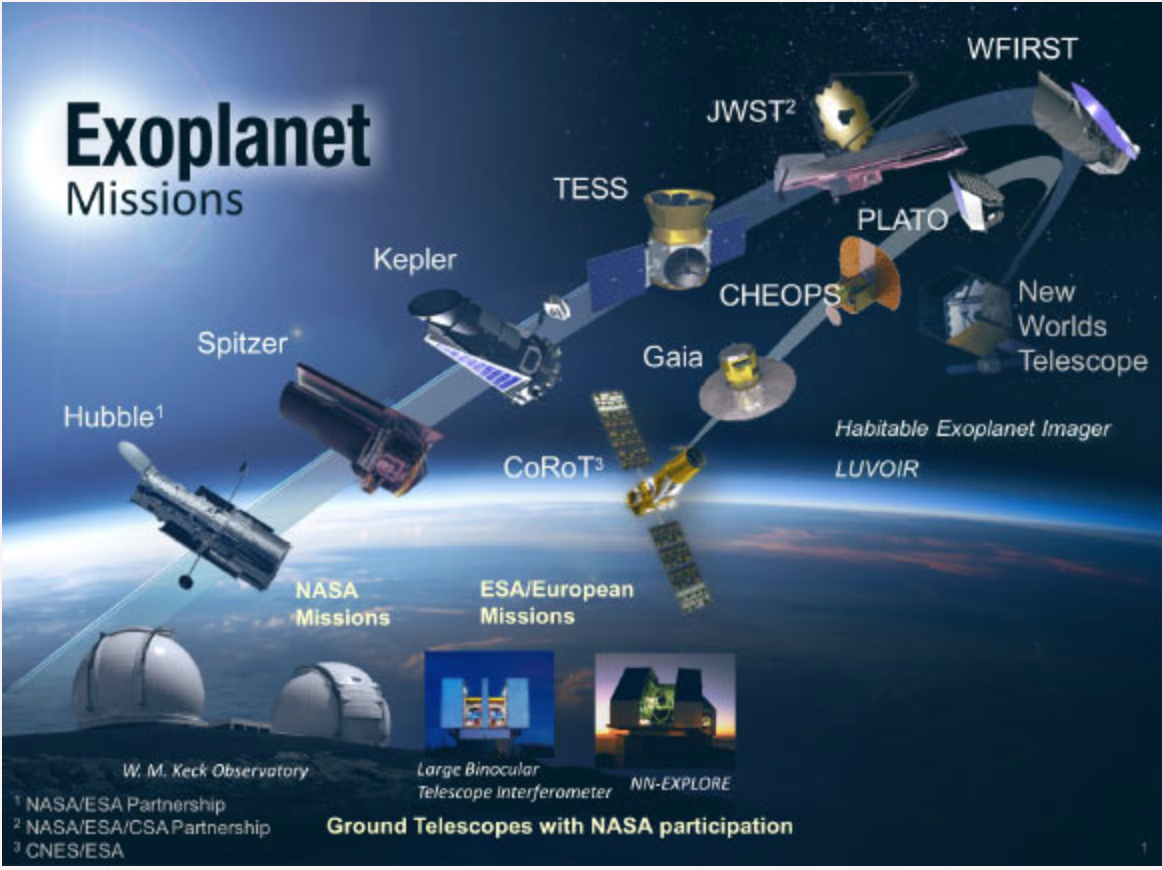}
 \caption{A schematic timeline of NASA and ESA exoplanet related space missions and the ground-based follow-up telescopes NASA directly participates in.}
  \centering
\end{figure*}

As new exoplanet transit missions such as K2 \citep{K2-2014}, and the currently operating missions TESS \citep{Ricker2015} and CHEOPS \citep{Benz2020} come along, follow-up high-resolution (sub-arcsecond) imaging continues to be needed and in larger amounts than before. 
While Gaia can resolve companions down to near 1.0 arcsec and a bit closer using additional observations over time, e.g., EDR3; Fabricius et al. arXiv:2012.06242), it does not reach the spatial resolution of speckle imaging. Additionally, other exoplanet search techniques such as radial velocity \citep{Kane2019} and ground-based small telescope transit surveys \citep{Bakos2007} also benefit from speckle imaging of any candidate systems. Finally, the next wave of exoplanet space telescope missions will soon be upon us (Figure 1); missions covering larger and deeper sky areas such as PLATO (transits), and those hoping to obtain detailed exoplanet science such as the James Webb Space Telescope (JWST; transit spectroscopy, emission spectroscopy, direct imaging of exoplanets) and the Nancy Grace Roman Space Telescope (formally the Wide Field Infrared Survey Telescope [WFIRST]; direct imaging and microlensing planets), as well as complete spectroscopic characterization of exoplanet atmospheres with Ariel (Figure 2). Anywhere high-resolution imaging is needed, including for future missions such as LUVOIR, HabEX, or OST, our speckle program will be valuable. By that time, it is hoped that speckle imaging will be an integral part of the 30-m ground-based telescope system, providing angular resolutions near 5 mas.

For JWST, our program will provide high resolution imaging in support of targeted exoplanets and their host stars. Roman will make use of speckle imaging to support exoplanet research in two main ways: First, to vet and fully characterize direct exoplanet imaging targets in order to assess their multiplicity, and secondly, imaging of microlens sources to aid in the characterization of the source and lens stars.

\begin{figure}
\centering
\includegraphics[scale=0.25,keepaspectratio=true]{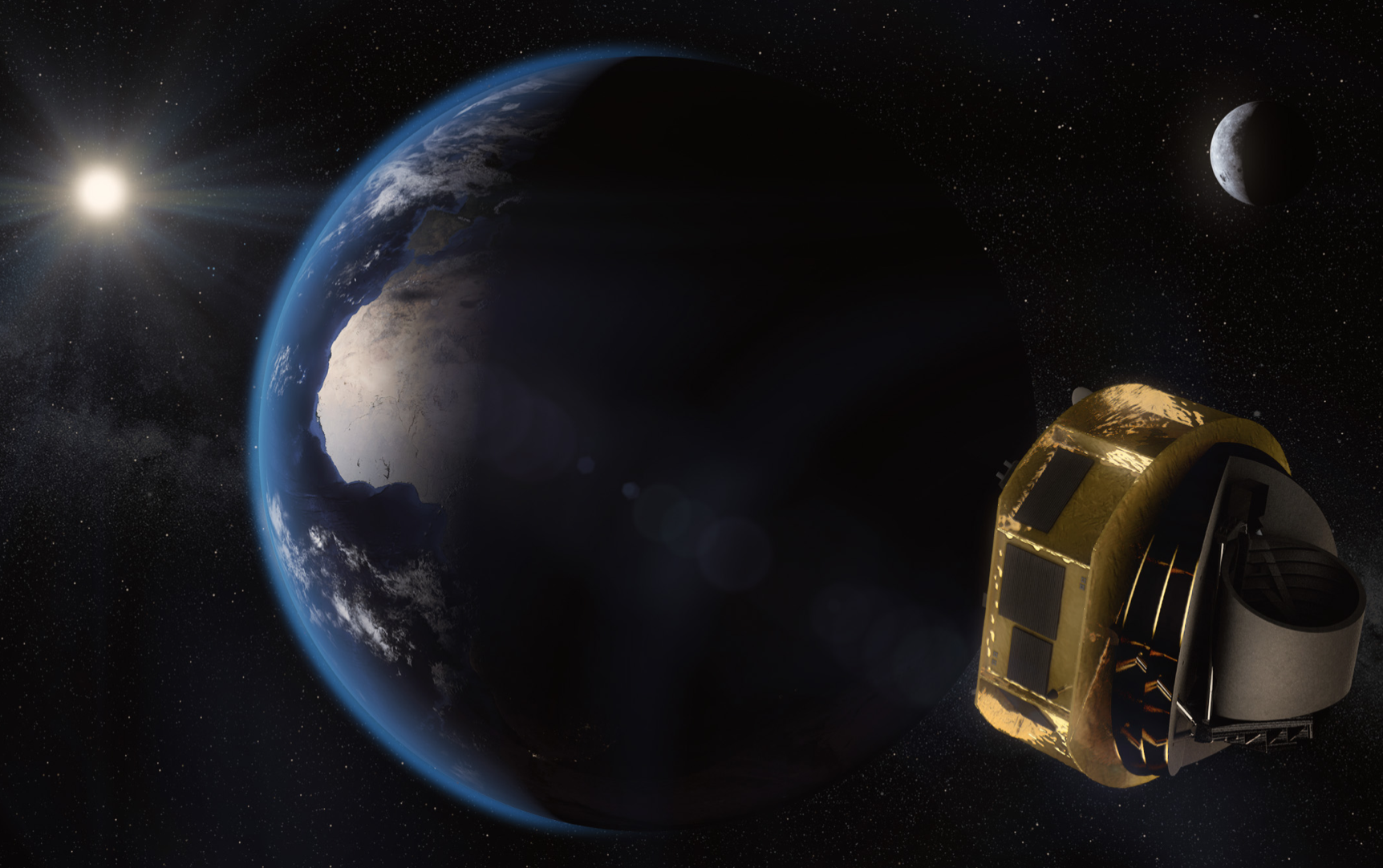}
 \caption{Artist concept of the Ariel space telescope.}
  \centering
\end{figure}

This paper provides an overview of the
NASA high resolution speckle imaging program.
Exoplanet transit and radial velocity studies mainly focus on (A)F to M stars, however, our speckle imaging techniques have been used for research programs related to stars of all spectral and luminosity classes, extended objects, and solar system bodies. These applications are not discussed further in this report.
Section 2 discusses the need for high-resolution imaging, \S3 presents the NASA mechanism to engage the exoplanet community, \S4 and \S5 give an overview of the instrumentation used, the community program, and data produced in this program, \S6 lists some of the major scientific discoveries the speckle program has made in relation to exoplanet host star multiplicity, and finally we summarize in \S7.

\section{The Need for High-resolution Imaging}

Survey telescopes, such as Kepler and TESS, cover a wide field of view, but have large pixels on the sky. Kepler (and K2) had 4 arcsec/pixel values in their focal plane and TESS has 20 arcsec/pixel. These large pixels gather all the light from any stars present within the extracted photometric apertures. If a transit-like event is detected, it is not immediately obvious which star in the pixel (or actually in the pixels) used for light curve construction is the cause of the event. Thus, the status as a real exoplanet transit candidate remains in question until some form of validation is carried out.

Telescopes such as Hubble have great spatial resolution, but they come at the cost of a small field of view and large over-subscription rates for observational proposals. While space has the advantage of stable observing conditions and no atmospheric effects, high-resolution imaging from the ground must make use of clever means to attempt to ``remove" the blurring effects of the atmosphere. IR/AO uses (laser) guide stars and deformable mirrors while speckle interferometry freezes the atmospheric distortions using many short exposures and reconstructs these into diffraction limited images using specialized software techniques.

Figure 3 illustrates these points for the case of KOI-1002 imaged from a typical ground-based telescope, Kepler, and TESS. The two bright stars in the top row of the figure are approximately 25 arcsec apart. The two stars are still separate in the Kepler image but near enough to place scattered light or even be both captured in any aperture used to measure the photometry. In the TESS image, it is difficult to even understand the scene. A good ground-based image, such as the top left image, can be used to help understand the star field imaged by TESS, but does not answer the question of which star had the transit-like event. 
The bottom row of Figure 3 shows optical speckle results obtained at Gemini and an IR adaptive optics image obtained at Keck for the star at the center of the top panel. Note the factor of $\sim$50 scale change in spatial dimension. 
We see that the speckle and the IR/AO images reveal that the central star is actually a binary system, thus while the transit-like event could be real, the properties of the planet and its host star would need revision due to the third-light of the close, likely bound companion.

\begin{figure*}[t]
\centering
\includegraphics[scale=0.45,keepaspectratio=true]{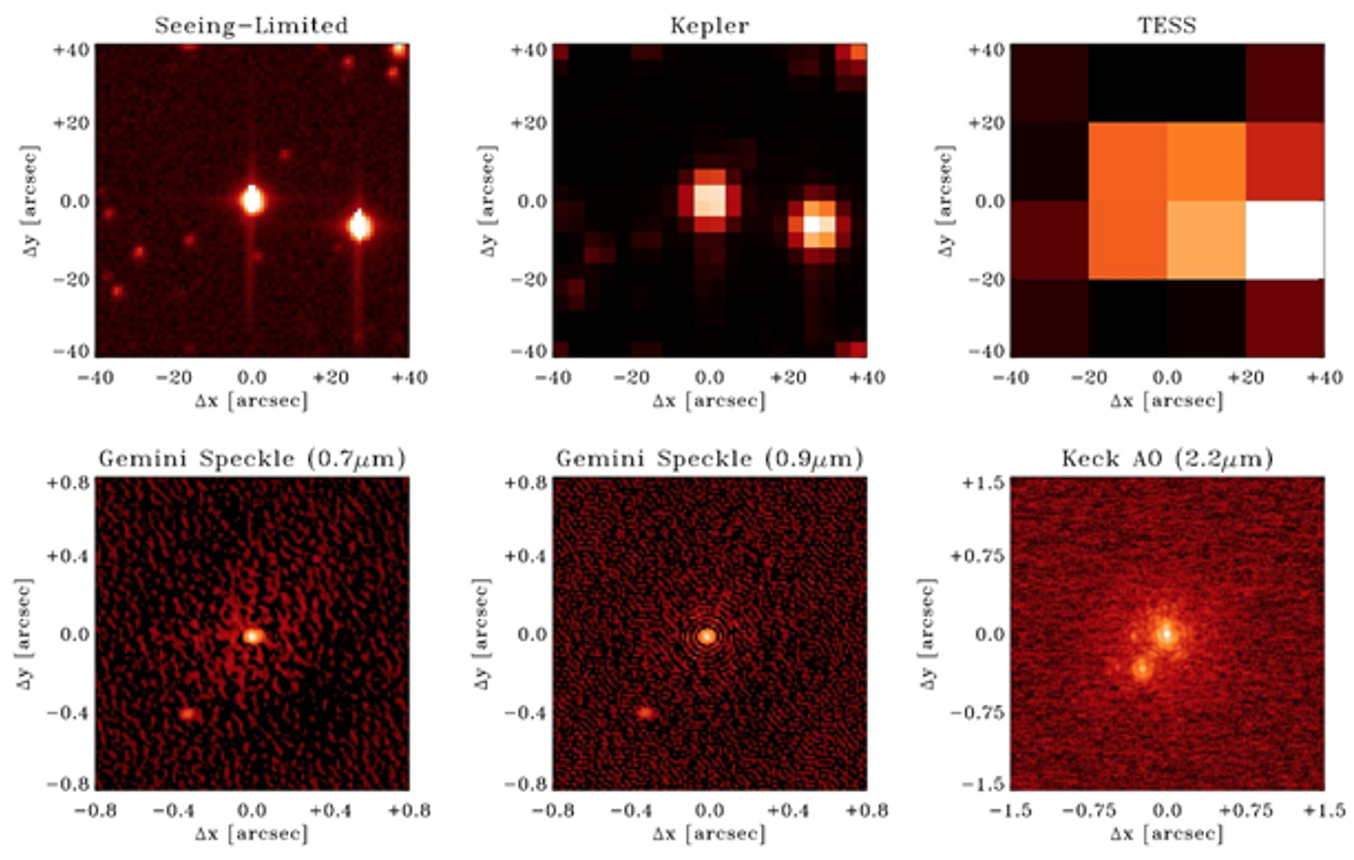}
 \caption{(Top row) Ground-based seeing-limited image of KOI-1002 (star at center) obtained from the Palomar Observatory Sky Survey (POSS). The image shows the local star field which contains two bright stars.
 The Kepler image is from the standard 30-minute postage stamp images downloaded from the spacecraft and the TESS ``image" was produced (before TESS had observed the Kepler FOV) by convolving the Kepler image with the TESS 20 arcsec pixel size kernel. (Bottom row) Here the much smaller spatial dimension boxes are centered on the bright central star in the top row and it is revealed in the high-resolution images that the star is a binary system with a separation of about 0.5 arcsec. The Keck AO image shows the central 3 arcsec box. The speckle images took a total time of 5 minutes (3 minutes on source) and reached a contrast of 6.5 magnitudes while the IR/AO image required 15 minutes total time ($\sim$1.5 minutes on source) and reached a contrast limit of 9.2 magnitudes. The inner working angle for the speckle and IR/AO observations was $\sim$20 mas and $\sim$50 mas respectively.}
  \centering
\end{figure*}

These types of follow-up high-resolution images are key to understanding the light within the scene of an exoplanet host star candidate. This is an important step in exoplanet validation and characterization.
If the star is indeed multiple, and we know that about 40-50\% of exoplanet host stars have one or more stellar companions \citep{Horch2014,Matson2018}, then knowledge of the brightness and type of any companion stars are crucial in order to properly assess the exoplanet and host star properties. 

\citet{Ciardi2015,Wang2015ApJ...813..130W,FH2017,Deacon2016MNRAS.455.4212D} and \citet{ Zeigler2018AJ....156...83Z}, for example, have shown that the presence of third-light will mean that the planet radius determined from the transit depth alone is incorrect, the planet will always be larger than estimated from the transit depth, at times so large as to lose planet status. \citet{FH2017} noted that such third-light properties will also decrease the mean density of the planet, possibly turning a terrestrial exoplanet into an ice giant as well as causing atmospheric scale height calculations to be flawed. These same two authors \citep{FH2020} also showed how the lack of knowledge 
of a companion star could cause measured stellar properties, such as metal content and log g, to be incorrectly derived from an analysis of the star's spectrum. Use of the knowledge of a companion (or not) allows a proper characterization of both the exoplanet and stellar properties. High resolution knowledge of the scene around host stars will remain an important diagnostic for future transit, direct imaging, microlens, and atmospheric spectroscopy exoplanet missions.

As an aside, ROBO-AO is another high-resolution imaging technique used in the optical wavelength range. \citet{Zeigler2017AJ....153...66Z} discuss their results using this method for exoplanet host stars. Unlike speckle imaging, ROBO-AO uses the mechanical deformable mirror techniques of IR/AO and applies them to optical light. To date, this application has suffered from the use of small aperture telescopes ($<$2-m) at sites providing modest native seeing ($\sim$1 arcsec) and only low-order AO corrections due to the very short wavefront coherence times available in the optical. As such, ROBO-AO
observations resolve roughly equal mass binaries at separations $\geq$0.8 arcsec with an increasing delta magnitude contrast from $\sim$1 at 0.9 arcsec to $\sim$3 at 1.4 arcsec.

\section{The Speckle Imaging Program}

Our speckle imaging program is set up under the auspices of NASA through the Exoplanet Exploration Office (ExEP) NN-EXPLORE program located at the Jet Propulsion Laboratory (JPL). The ExEP provides advocacy for the exoplanet community to NASA in terms of future mission science directions, needed technology, and exoplanet science critical to enable a full understanding of exoplanets, their environments, and the search for life. As an aid to establishing an open forum with the community, the ExEP maintains two ``gap lists", one for technology\footnote{https://exoplanets.nasa.gov/exep/technology/gap-lists/} and one for science\footnote{https://exoplanets.nasa.gov/internal\_resources/1547/}. Each of these documents represent community vetted gaps, that is, areas which need additional understanding in the pursuit of exoplanet science.
The ExEP science gap list contains twelve specific areas that the exoplanet community has agreed are in need of further, detailed understanding. Our speckle interferometry program directly addresses five of these science gaps and enables four additional ones. 

\bigskip

~~~~Directly Addresses

\begin{itemize}
\item Science Gap 12 - Measurements of Accurate Transiting Planet Radii

Speckle imaging provides knowledge of companions, especially true bound companions and, if detected, the ability to correct the exoplanet radius and other properties for ``third light".

\item Science Gap 07 - Properties of Known Exoplanet Host Stars

Speckle imaging provides knowledge of the multiplicity of exoplanet host stars providing accurate stellar parameters and directly assessing the topics of exoplanet formation, migration, dynamics, and evolution. 

\item Science Gap 10 - Precursor Observations of Direct Imaging Targets

Speckle imaging provides exploratory observations of potential targets for future direct imaging and atmospheric observation missions, assessing their multiplicity and thus their potential as high-value targets. 

\item Science Gap 04 - Planetary System Architectures: Occurrence Rates for Exoplanets of all sizes

Speckle imaging allows the correct exoplanet and stellar properties to be determined. This in turn is used to derive robust occurrence rates for exoplanets orbiting stars in multiple star systems.

\item Science Gap 05 - Occurrence Rates and Uncertainties for Temperate Rocky Planets (eta-Earth)

Speckle imaging addresses exoplanet occurrence rates as described above and allows habitable zone locations in binary host star systems to be determined yielding eta-Earth rates for temperate planets.

\bigskip
\bigskip

Enables

\item Science Gap 01 - Spectral characterization of atmospheres of small exoplanets 

\item Science Gap 02 - Modeling exoplanet atmospheres 

\item Science Gap 03 - Spectral signature retrieval 

\item Science Gap 06 - Yield estimation for exoplanet direct imaging missions

\end{itemize}

\section{The Speckle Instruments}

Just before the Kepler mission was launched, we began our speckle imaging work to support the NASA community. At that time, we used the Differential Speckle Survey Instrument \citep[DSSI;][]{Horch2009} on the 3.5-m WIYN telescope located on Kitt Peak in southern Arizona, USA. This instrument was a workhorse during the Kepler mission,
providing high-resolution images of $\sim$1000 Kepler Objects of Interest (KOIs). Since that time, three new instruments of similar overall design but with increased functionality, larger and faster EMCCD imagers, and modernization in terms of automation, user interface, filter wheels, and remote operation have been built and deployed. The three new instruments are the NN-EXPLORE Exoplanet Star and Survey Imager \citep[NESSI\footnote{https://www.wiyn.org/Instruments/wiynnessi.html};][]{Scott2018} at the 3.5-m WIYN telescope, and 'Alopeke and Zorro\footnote{https://www.gemini.edu/instrumentation/current-instruments/alopeke-zorro}, (Scott et al. 2021, in prep.) duplicate instruments at the twin Gemini 8-m telescopes in Hawaii and Chile.

Each of the new instruments provide simultaneous observations in two optical bands, determined by filters placed in each of two beams, split at 700 nm. Figure 4 presents a schematic of the Gemini instruments with the major parts labeled and Figure 5 shows one of the instruments being constructed in our optics lab at NASA Ames (left) and the completed instrument mounted on the Gemini telescope (middle and right). Table 1 presents the general parameters of these new instruments with 'Alopeke and Zorro being identical therefore having identical parameters. The instruments have two field of view options user selectable in real-time; a narrow speckle imaging field and a wider more traditional imaging field of view. The angular resolution of these instruments provide inner working angle spatial resolutions for nearby exoplanet host stars (e.g., TESS) of $<$10 au \citep{Matson2019}.

These instruments are fully integrated into the telescope control systems where they reside and the Gemini instruments are permanently mounted for use at any time during the year. `Alopeke and Zorro are able to be operated remotely from the Gemini control room or anywhere internet is available via a secure connection to the observatory.

\begin{deluxetable*}{cccccccc}
\centering
\tablecaption{Speckle Interferometers} 
\tablehead{\colhead {Instrument} &
\colhead {Telescope} & \colhead{Detectors\tablenotemark{a}} & 
\colhead{FOV(")\tablenotemark{b}} & 
\colhead{"/pixel\tablenotemark{b}} &
\colhead{Filters\tablenotemark{c}} &
\colhead{Resolution (mas)} &
\colhead{Limiting Mag (R)}
}
\startdata
NESSI & WIYN & EMCCD & 19/83 & 0.02/0.08 & SDSS ugriz + 4 narrow band & 39 @550nm/ 64 @880nm & $\sim$13.2 \\
'Alopeke & Gemini-North & EMCCD & 6.7/60 & 0.01/0.1 &SDSS ugriz + 4 narrow band & 17 @562nm/ 28 @832nm & $\sim$18 \\ 
Zorro & Gemini-South & EMCCD & 6.7/60 & 0.01/0.1 & SDSS ugriz + 4 narrow band & 17 @562nm/ 28 @832nm & $\sim$18 \\ 
\enddata
\tablenotetext{a}{Each instrument uses two Andor iXon Ultra 888 back-illuminated Electron Multiplying (1024X1024, 13 micron) pixel CCDs (EMCCDs).}
\tablenotetext{b}{Values given are for the ``speckle" / ``wide-field" of view modes.}
\tablenotetext{c}{Narrow band filters are centered at 466nm, 562nm, 716nm, and 832nm.}
\centering
\end{deluxetable*}

\begin{figure}[h]
\centering
\includegraphics[scale=0.48,keepaspectratio=true]{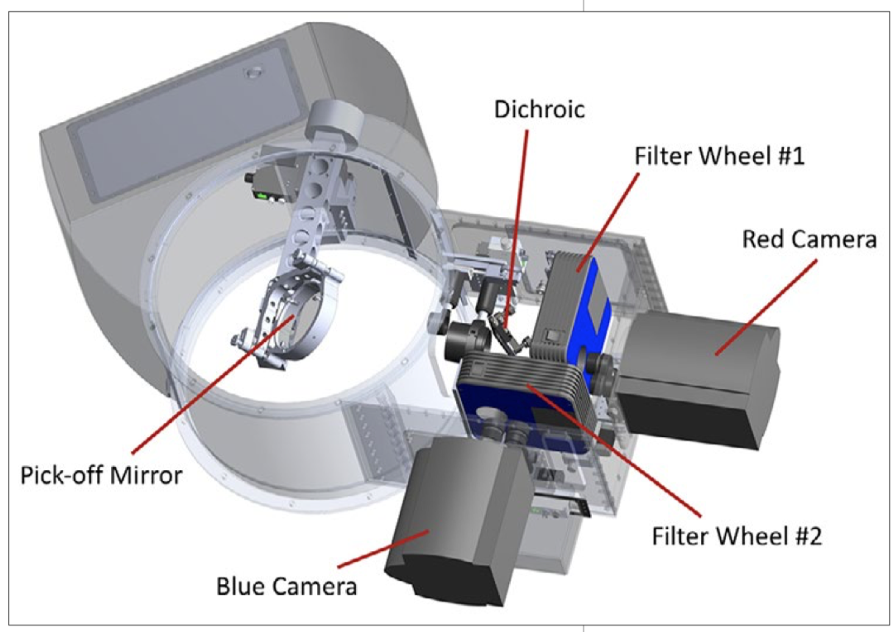}
 \caption{Schematic CAD drawing of 'Alopeke or Zorro (they are identical), the speckle instruments residing at the twin Gemini 8-m telescopes. The compact design is illustrated along with the primary parts labeled. 'Alopeke and Zorro both mean fox in the indigenous language of the local people.}
  \centering
\end{figure}

\begin{figure*}
\centering
\includegraphics[scale=0.55,keepaspectratio=true]{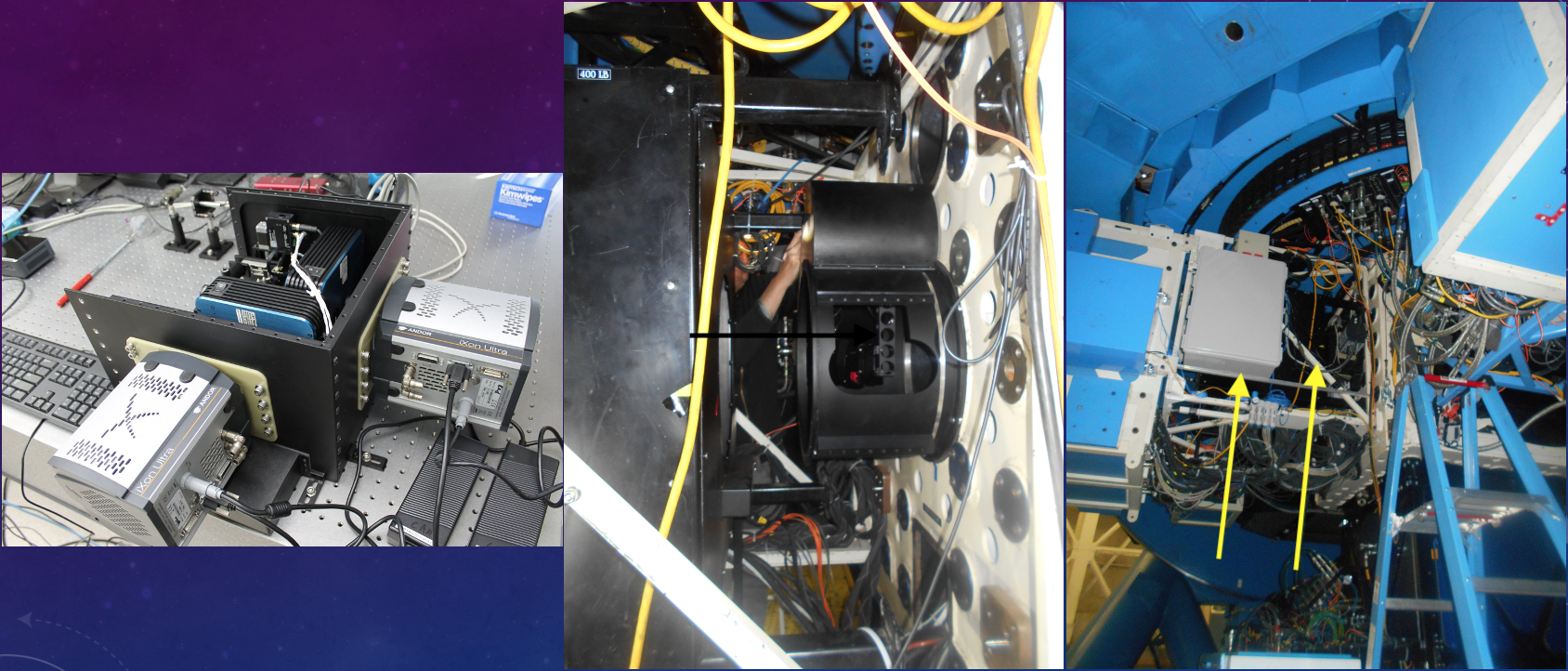}
 \caption{(Left) One of the two identical speckle imagers under construction at the NASA Ames Research Center. The two grey boxes extending out of the black box are the two Andor EMCCD cameras. (Middle) A close-up view of the instrument mounted on Gemini at the GCAL port and
 (Right) a view of the instrument (black box) with the associated power supply, electronics, and computer unit (white box). See yellow arrows.}
  \centering
  \bigskip
\end{figure*}

\section{The Community Observing Program}

The NASA speckle imaging program is openly available to the world-wide exoplanet community. While we have built and deployed the speckle instruments, our observing protocol is a community-based program with targets and observing priorities set by the missions and the community. 

Below, we discuss the observational program, the target selection methodology, the data reduction processes, and the archives which house the raw and reduced data products.

\subsection{Observations}
Proposals to use the speckle instruments, including proposals submitted by our team, are peer reviewed for each telescope by the relevant telescope allocation committee (TAC). Once the approved programs are known for a semester, we work with each observatory to set up block scheduled observation runs, once or twice each semester on each telescope. Each run consists of 6-10 nights depending on time demand.

All speckle targets are placed in a queue to be observed by our team during the observing block. Most targets are observed in the usual manner \citep{Howell2011} that is, thousands of 40 to 60 msec images are simultaneously collected in two narrow band filters, one each in the blue and red regions of the optical bandpass.
Here blue and red are defined by the dichroic in our instruments, splitting the optical light at 700 nm (See Table 1). Some observations may come with special requirements such as the use of specific filters, extended integration times, or specific time constraints.

\subsection{Target Selection}
Targets to observe are selected in three main ways. First, each space mission or ground-based exoplanet program has a team of scientists and staff designated to produce a priority list of targets to be observed. These targets, usually called ``objects of interest" (e.g., Kepler objects of interest, KOIs or TESS objects of interest, TOIs) are listed in priority order and provided to our observing team ahead of each observation run. 
For example, the TESS Follow-up Observing Program Working Group as well as the TESS sub-group for high-resolution imaging (SG3) play roles in target ranking and selection. 
The priority order reflects target brightness, number, location or type of exoplanet candidates, or stellar properties. The list contains 100-200 targets at a time and gets updated throughout each observing season. 

The second main way is direct communication with our group. If an exoplanet host star is in need of high-resolution imaging, regardless of what space mission or ground-based telescope discovered it or who is currently studying the target, we will add it into our priority list and observe the target for the PI. This method usually involves 1 to a few targets of immediate interest to someone in the exoplanet community.

Finally, PIs can propose for telescope time themselves as each of our instruments at their associated telescopes have ``open sky" policies, that is, proposals are accepted from anyone. 

\subsection{Data Reduction}

Speckle interferometry produces diffraction-limited images over a small field of view ($\sim$1 arcsec) by using very short  exposure times to "freeze" the atmosphere. Fourier or other techniques are then used to reconstruct these images and search for interference fringes that exist if two or more stars are within the narrow field of view (See Figure 6). 

\begin{figure}[h]
\centering
\includegraphics[scale=0.50,keepaspectratio=true]{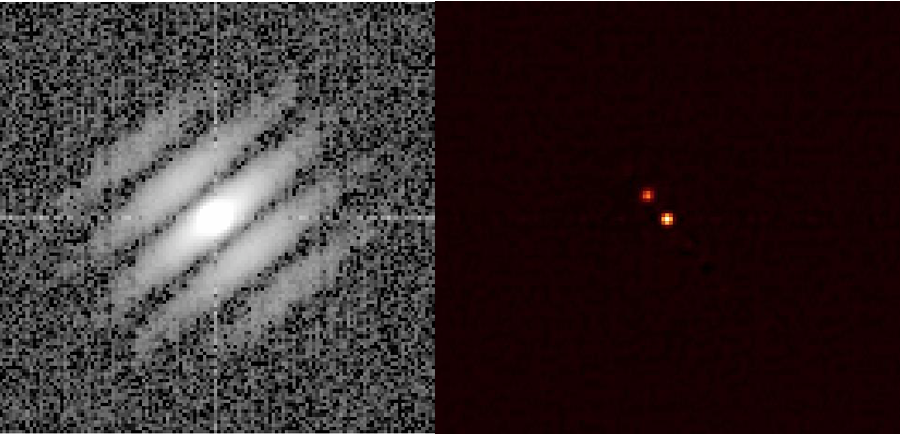}
 \caption{Power spectrum of TOI-1356, an exoplanet hosting star, produced by our reduction pipeline using 2000 speckle images and obtained at the WIYN telescope. Note the interference fringes whose spacing, intensity, and orientation allow a reconstructed image (right) to be produced. From this image and the Fourier analysis, we can determine the separation, position angle, and magnitude difference of the two stars. In this case, the host star is a member of a close binary having a separation near 0.16 arcsec. The reconstructed image is 2 arcsec on a side.}
  \centering
\end{figure}

Speckle imaging has been used since the early 1970's but has been greatly upgraded since then, now using digital detectors, large telescopes, and advanced software techniques.
For example, EMCCDs provide for electron multiplication in the output gain register, increasing the input signal by up to 1000 times. 
Stars as faint as 19th magnitude can now be observed using our instruments at Gemini coupled with their EMCCD detectors.  

Our standard data reduction pipeline and the data products available are described in \citet{Howell2011} with some additional reduced data products now provided (e.g., Figure 7). The following reduced data products are delivered to each PI: sensitivity curves, reconstructed images, and any binary fit parameters if a companion is detected.

\begin{figure}[h]
\centering
\includegraphics[scale=0.42,keepaspectratio=true]{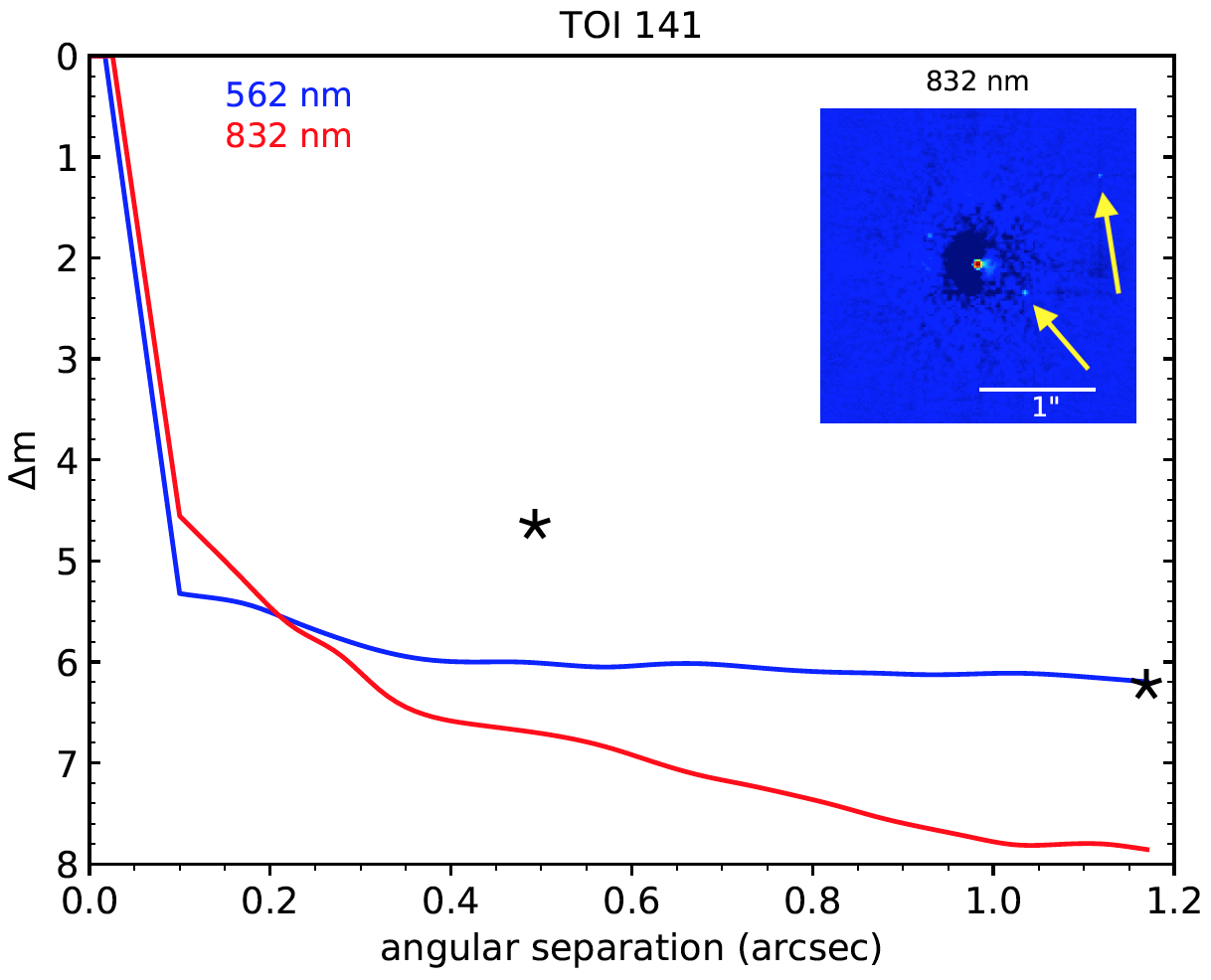}
 \caption{One of our standard pipeline reduced data products. The plot shows our 5$\sigma$ contrast curves in both filters as a function of the angular separation out to 1.2 arcsec, the end of speckle coherence.
 In addition, the inset shows the reconstructed 832 nm image with a 1 arcsec scale bar. This star, TOI 141, was found to have two close companions, one at 0.5 arcsec (PA=240 degrees, Delta magnitude = 4.6) and one at 1.3 arcsec (PA=307 degrees, Delta magnitude = 6.2). The yellow arrows mark the companions in the reconstructed image and the black stars mark the locations of the stars on the main plot, the 1.3 arcsec companion placed at 1.2" to fit within the plot.
 }
  \centering
\end{figure}

\subsection{Data Archives}

Our raw data and reduced data products are archived at both the Gemini data archive\footnote{https://archive.gemini.edu/searchform} and the NASA Exoplanet Archive, Exoplanet Follow-up Program (ExoFOP) archive\footnote{https://exoplanetarchive.ipac.caltech.edu}.
They are accessible to the public once the proprietary period has ended. All of our community observations, that is those observed within the NASA proposed program time, are available to the public with no proprietary period. Other PIs observing exoplanet targets can specify a proprietary period of up to 12 months, however, most PIs choose no proprietary period as well.

All observed targets are listed on the Exoplanet Follow-Up Observation Program website\footnote{https://exofop.ipac.caltech.edu/tess/} within days of an observing run; in this way the community is informed about which exoplanet host stars have been imaged with our speckle instruments. Final data products from the reductions are also posted on ExoFOP; tags link them to the corresponding entries in the imaging observations tables. For targets observed as part of our community program, the reduced data products are posted on ExoFOP soon after an observing run ends and the reduction has been completed, typically within a few weeks. To date, we have observed over 1,000 exoplanet host stars from Kepler and K2 and more than 500 TESS mission TOIs, so far, for the community. These observations have been used to validate and characterize exoplanets in over 65 published papers in the past 1.5 years. 

\section{Discoveries in Stellar Multiplicity}

In addition to validating and characterizing numerous exoplanet discoveries as contributions to the exoplanet community research, we have used the large samples of exoplanet host stars observed at high resolution to enable a number of overarching findings.

1) {\it{Percentage of binary exoplanet host stars:}} For both the Kepler and K2 exoplanet host stars, and it seems true also for TESS, we have shown that for those stars hosting at least one exoplanet, 40-50\% of them reside in binary or multiple star systems; See \citep{Horch2014,Matson2018}
and the upcoming papers by Howell et al. (2021) and Lester et al. (2021).

2) {\it{Binary host star properties:}}
Recent work (Howell et al. 2021) has shown that the mass ratio of exoplanet host star binaries follows that of field binaries, that is an excess of nearly equal mass systems. The orbital period distribution however does not. These authors find that exoplanet hosting binary stars show generally larger mean separations having a peak in their orbital separation near 100 au, not 40 au as for the field sample \citep{Rag2010}.

\citet{FH2017,FH2020} have shown that it is critical to fully understand the ``scene" of light near an exoplanet host star. If a close companion is present, both the exoplanet and the stellar properties determined may be in error.

3) {\it{Bound vs. line of sight companions:}} Using statistical modeling, \citet{Horch2014} and \citet{Matson2018} have shown that most stellar companions  that reside within 0.4 arcsec (at Gemini) and 0.8 arcsec (at WIYN) of the primary star are true bound companions (at $\ge$90\% probability). These limits are important to note as stellar companions found to lie farther from the host star are most likely, but not exclusively, line of sigh companions, therefore not part of the formation, dynamical, and evolutionary processes of the exoplanetary system. However, they nonetheless cause transit dilution, and so their flux has to be taken into account when the transit depth and exoplanet properties are derived.
Observational techniques such as lucky imaging or ROBO-AO tend to only detect companion stars which are 0.75 arcsec or farther away. Even Gaia cuts off at direct companion detection near 1.0 arcsec. At such sub-arcsecond angular resolutions, speckle imaging provides inner working angles for many of the target stars of 1 to a few au.

\citet{Everett2015} and \citet{Hirsch2017} also used model isochrones to provide evidence that companion stars around KOIs are bound or not. Finally, \citet{Colton2020} has begun to measure space and orbital motions for exoplanet host stars with stellar companions, a process that takes decades for the far away Kepler stars, but will only take a few years for some very close K2 and TESS binary host stars with separations of $\le$10 au. These astrometric measurements have already begun to provide information on the formation, dynamics, and evolution of exoplanet systems.

4) {\it{Which star in a binary hosts the exoplanet:}}
Transit discoveries in binary host stars often leave open the question of which of the stars the exoplanet(s) actually orbit. If the planet orbits the primary, the transit depth may require only a modest correction due to a fainter secondary star. However, if the transit is due to something orbiting the secondary star or the two stars are nearly equal in brightness, it can be very unclear what the measured transit depth is really telling us. One way to solve this dilemma was employed by  
\citet{2019AJ....158..113H}
for the binary A star exoplanet host Kepler-13.
Using simultaneous time-series speckle observations of both
stars in the pair, it was shown that the transit occurs on Kepler-13A.

\section{Summary}

We have summarized our continuing decade long NASA high resolution imaging work for exoplanet research. Using speckle interferometry, we carry out a community led observational program that supports space- and ground-based exoplanet efforts. Observations are obtained at the WIYN 3.5-m telescope in Arizona, and at both the Gemini-North and Gemini-South 8-m telescopes located in Chile and Hawaii. We have designed and built new instruments for these telescopes that are available to the community through peer review proposals under the NN-EXPLORE/NOIRLabs open-sky policies. All our observations and their fully reduced data products are made available via public data archives.

Sub-arcsec imaging, especially inside of 0.4 arcsec, is critical 
for our detailed understanding 
of exoplanets, their host stars, and the search for other life in the Universe. Exoplanet radii and mean densities, plus the stellar properties of their hosts, can be incorrectly determined without proper knowledge of the close-in light scene. 

Our imaging program has supported
many exoplanet validation and characterization studies for space missions and ground-based surveys, and radial velocity studies as well as made scientific findings itself along the way.

We plan to continue our community service program throughout the TESS extended mission and into the JWST and Roman missions and beyond.

\section*{Conflict of Interest}
The authors declare that the discussion and research was conducted in the absence of any commercial or financial relationships that could be construed as a potential conflict of interest.

\section*{Acknowledgments}
We thank
Elliott Horch and William Sherry for their previous team work. Additionally, we thank the generous support and collaboration of the staff at the WIYN and Gemini Telescopes, they are indeed part of the Team. We appreciate the reviews by the referees which led to a better paper. Finally, we'd like to thank NASA headquarters and the Exoplanet Program Office at JPL, in particular, Doug Hudgins, Gary Blackwood, and John Callas, for their substantial support of the speckle imaging program over these many years. 

{\it Facilities:} WIYN:3.5m (NESSI), Gemini-North:8-m ('Alopeke), Gemini-South:8-m (Zorro)

\newpage
\bibliographystyle{apalike}
\bibliography{MAIN}

\end{document}